\documentclass{iopart}
\usepackage{graphicx,stackrel}
\newcommand{\be}{\begin{equation}}
\newcommand{\ee}{\end{equation}}

\usepackage{amssymb,color}

\begin{document}

\title{Electronic transport in three-terminal chaotic systems with a tunnel barrier} 

\author{Lucas H. Oliveira$^1$, Anderson L. R. Barbosa$^2$, Marcel Novaes$^1$}
\address{$^1$Instituto de Física, Universidade Federal de Uberlândia, 38408-100, Brazil}
\address{$^2$Departamento de F\'{\i}sica, Universidade Federal Rural de Pernambuco, 52171-900, Brazil}

\begin{abstract}
We consider the problem of electronic quantum transport through ballistic mesoscopic systems with chaotic dynamics, connected to a three-terminal architecture in which one of the terminals has a tunnel barrier. Using a semiclassical approximation based on matrix integrals, we calculate several transport statistics, such as average and variance of conductance, average shot-noise power, among others, that give access to the extreme quantum regime (small channel numbers in the terminal) for broken and intact time-reversal symmetry, which the traditional random matrix approach does not access. As an application, we treat the dephasing regime.

\end{abstract}

\maketitle

\section{Introduction}

Electronic transport through mesoscopic samples has been an intense research topic, both theoretical and experimental, for the last three decades \cite{Altshuler,Beenakker,Alhassid,Mello,nazarov}.
Interest in the mesoscopic regime stems from the fact that electrons can maintain phase coherence throughout the process, leading to observable quantum effects.
One of the most important examples of ballistic mesoscopic sample are the chaotic semiconductor billiards, one of the platforms for studying quantum chaos \cite{Chandramouli, ChenRong, ALR, nathan,Bereczuk}, the interplay between quantum properties and  chaotic dynamics \cite{Eric,Hong,haake}.

One traditional tool to study quantum chaotic transport is Random Matrix Theory (RMT) \cite{Beenakker,Mitchell}, a statistical approach where operators are replaced by random matrices and which is well suited to reveal universal characteristics that are independent of geometric details of the system. Still, results are strongly affected by the intrinsic symmetries of the corresponding Hamiltonian, such as time-reversal symmetry (TRS) and spin-rotation symmetry \cite{Beenakker,Mitchell}. 
It has three basic ensembles: orthogonal ensemble, which preserves TRS and spin-rotation symmetry; unitary ensemble, which has TRS broken by a magnetic field; symplectic ensemble, which keeps TRS but has spin-rotation symmetry broken by spin-orbit interaction. 
The method can be applied directly to the Landauer-B\"uttiker approach \cite{landauer,buttiker} to calculate the expected value of electronic transport moments, such as conductance   \cite{Baranger,BB,Heusler,whitney,Waltner,kumar}, shot-noise power \cite{Blanter,Braun,Jalabert,whitney,anderson,anderson2,bar1,Waltner,whitney2,savin2,kumar2}, and the third cumulant \cite{Reulet,bar1,anderson3,savin1}.
Besides, in the limit of a large number of channels in the terminals, both the variance of conductance and average of shot-noise power hold universal value; those are a footprint of quantum chaos \cite{Beenakker}.

The Landauer-B\"uttiker approach relies on the scattering matrix $S$, which describes electronic transport in the presence of three terminals,
\begin{eqnarray}
{S}=
\left[\begin{array}{cccc}
\mathbf{r}_{11}&\mathbf{t}_{12}&\mathbf{t}_{13}\\
\mathbf{t}_{21}&\mathbf{r}_{22}&\mathbf{t}_{23}\\
\mathbf{t}_{31}&\mathbf{t}_{32}&\mathbf{r}_{33}\\
\end{array}\right],\label{M}
\end{eqnarray}
where $\mathbf{t}_{ba}$ is the $N_a \times N_b$ transmission matrix block from terminal $a$ to terminal $b$, while $\mathbf{r}_{aa}$ is the $N_a \times N_a$ reflection matrix block. Given two terminals $a,b$, we define the dimensionless moments that characterize transport from $a$ to $b$, the polynomials $p_\lambda$, as functions of transmission matrix $T = \mathbf{t}_{ab}^\dag \mathbf{t}_{ab}$, as follows. Given a sequence of positive integers $\lambda=(\lambda_1,...,\lambda_\ell)$,
\begin{equation}
p_\lambda(T)=\prod_{k=1}^{\ell}{\rm Tr}(T^{\lambda_k}).
\end{equation}
The observables of the theory are the average values of these spectral statistics, $\langle p_\lambda\rangle$, where the average is taken over a local energy window. For example, $\langle p_{(1)}\rangle$ is the average conductance, $\langle p_{(1,1)}\rangle-\langle p_{(1)}\rangle^2$ is the conductance variance and $\langle p_{(1)}\rangle-\langle p_{(2)}\rangle$ is the average shot-noise.

Two ingredients that influence the expected value of electronic transport moments significantly are the tunneling barriers effect \cite{BB,anderson,anderson2,bar1,anderson3,Reulet,Waltner,whitney2,poisson2}, caused by the junction between the terminal and the billiard, and overall number of terminals \cite{whitney,baranger,brouwer,anderson4,Ramirez}. 
Some exact results are available for ideally connected multiterminals in the literature \cite{Adagideli}. 
On the other hand, works that consider both the tunneling barriers effect and multiterminals are, in general, perturbative calculations that are only valid in the regime of a large number of open channels \cite{whitney,anderson,anderson2}.
Hence, non-perturbative calculations involving the tunnel barriers and multiterminal are missing in the literature. Besides, recent numeric calculations have shed light on the importance of studying the extreme quantum limit  \cite{nathan} when the number of channels in the terminal is very small.

In this work, we take a step towards reconciling multiterminals with the presence of a tunnel barrier, at the extreme quantum limit, for both unitary and orthogonal symmetry classes. The barrier is characterized by a single tunneling probability $\Gamma$ or by its reflectivity $\gamma=1-\Gamma$, and it may be present in any of the terminals, i.e. either in the ones participating in the transport or in a third one (treating more than one barrier is still a challenge). In principle, different reflectivities might be associated with each channel in a terminal, e.g. $\{\gamma_1,\ldots,\gamma_{N_3}\}$ in terminal 3, and these would be eigenvalues of $\bar{S}\bar{S}^\dagger$, where $\bar{S}$ is the average scattering matrix, but we only treat the case when all reflectivities are equal, for simplicity.

We rely on a novel semiclassical approach based on matrix integrals \cite{matrix,trs}, thereby extending recent results for a two-terminal system with one tunnel barrier and TRS broken \cite{Pedro}. We obtain expressions for transport statistics which are power series in the barrier's reflectivity, with coefficients which are rational functions of the channel numbers.
As an application, we revisit a dephasing model introduced by B\"uttiker \cite{buttiker1}. 
We have checked that our results agree with numerical simulations done by sampling $10^5$ random scattering matrix in the presence of the tunnel barrier.
We confront our findings with results obtained via the diagrammatic method in the RMT framework, which are however limited to large channel numbers \cite{BB}.

The work is organized as follows: Section II reviews the semiclassical approximation and its efficient implementation using matrix integrals. In Section III, we present our general result  for $\langle p_\lambda\rangle$ with three terminals with one tunnel barrier and TRS breaking, which allows study of the average conductance, variance of conductance, average shot-noise power, and average of the third cumulant, in different limits. To finish this section, we use the RMT diagrammatic method to obtain results in the limit of large channel numbers to confront with our semiclassical approximation. In Section IV, we turn to the case of systems with intact TRS, for which we can compute the average conductance. In Section V, we apply the results of Sections III and IV to study the effect of dephasing in the presence of a tunnel barrier. Finally, conclusions can be found in Section VI.

\section{Semiclassical approximation}

\subsection{Encounters and diagrammatic rules}

The statistical moments characterizing transport from terminal $a$ to terminal $b$ can be written as 
\begin{equation}\label{moment}
p_\lambda(T)=\sum_{\vec{i},\vec{o}}\prod_{k}(\mathbf{t}_{ba})_{o_k,i_k}(\mathbf{t}_{ba}^\dagger)_{o_k,i_{\pi(k)}},
\end{equation}
where $\pi$ is a permutation with cycletype $\lambda$, i.e. having a cycle of length $\lambda_1$, a cycle of length $\lambda_2$, etc.

The semiclassical approximation to a matrix element $S_{ij}$ involves a sum over all possible trajectories starting at channel $j$ and ending at channel $i$. Each trajectory contributes a complex number whose argument is the classical action and whose modulus is related to the stability of the trajectory, i.e. its Lyapunov exponent. In the semiclassical expression for
(\ref{moment}) we end up with some trajectories coming from $\mathbf{t}_{ba}$ (\emph{direct} ones), going from $i_{k}$ to $o_{k}$, and
some other trajectories coming from $\mathbf{t}_{ba}^\dagger$ (\emph{partner} ones), going from $i_{\pi(k)}$ to $o_k$. When
a local energy average is considered, $\langle
p_\lambda(T)\rangle$, over an energy window which is small in the classical scale but large in
the quantum scale, the result vanishes as $\hbar\to 0$ unless there happens to be constructive interference betwen the direct and the partner trajectories, so that the total action is stationary. Under such a stationary phase approximation, the result
is determined by so called ``action correlations'': partner trajectories must have almost the same collective action as direct ones.

\begin{figure}
    \centering
    \includegraphics[scale = 0.35]{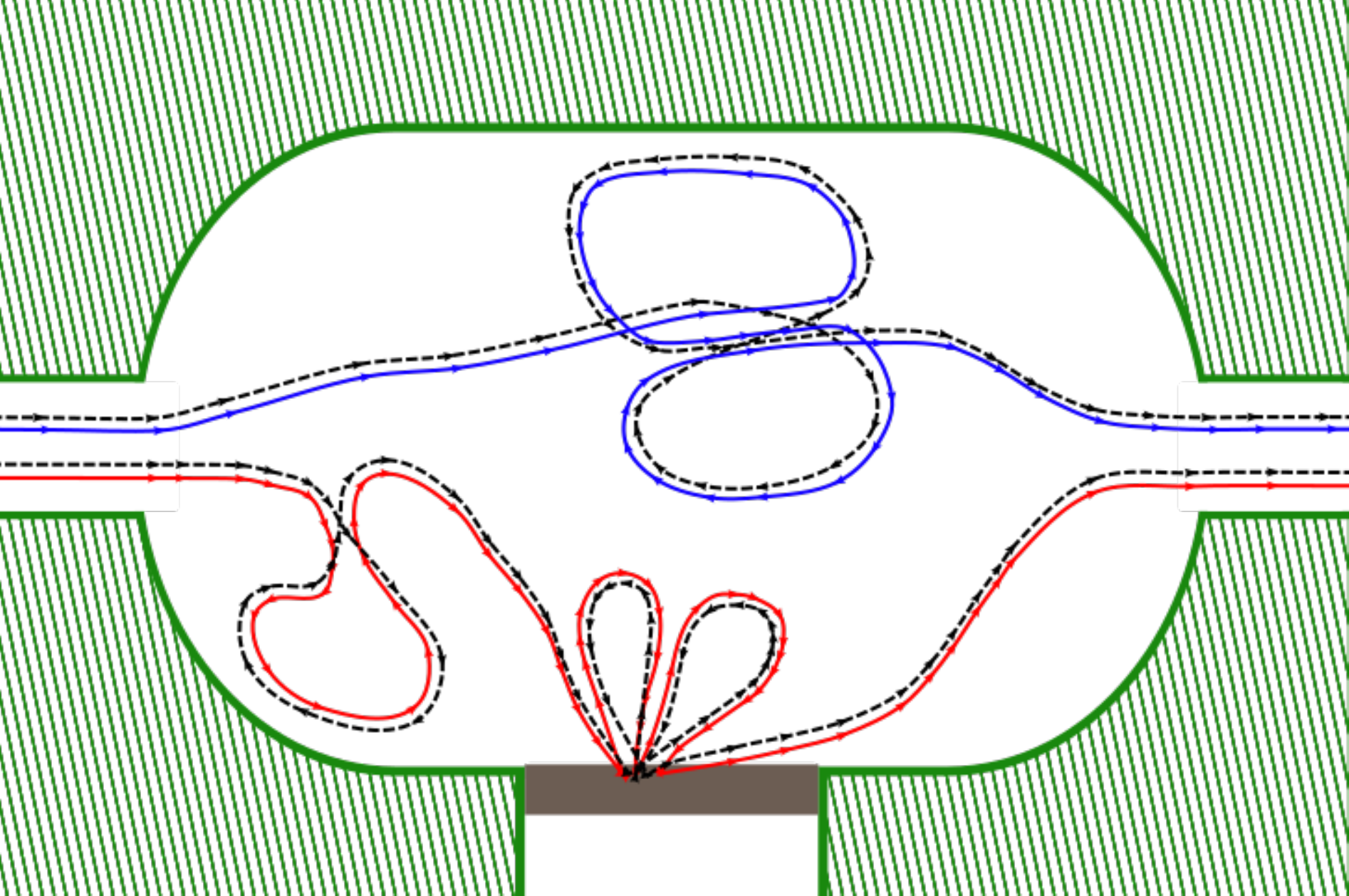}
    \caption{Chaotic mesoscopic billiard connected to three terminals having one of them as a tunnel barrier (grey). The lines illustrate the possible trajectories of particle through the billiard. }
    \label{trajetorias}
\end{figure}

This theory of correlated chaotic trajectories was breakthrough in semiclassical physics and has been discussed in detail in a number of papers \cite{sieber2,essen3,essen4,essen5,greg1,greg2,greg3}, to which we refer the reader for details. The set of partner trajectories differs from the set of direct ones only in very small regions (called \emph{encounters}).
A $q$-encounter is a region where $q$ pieces of trajectories run nearly parallel or anti-parallel (in the presence of time-reversal symmetry), and the difference between direct and partner trajectories is an effective permutation, i.e. they trace the same lines in space, but not in the same sequence. This ensures that they have almost the same collective action.

These trajectory multiplets are usually represented by diagrams, in which the encounters are represented by vertices and the complicated
pieces of chaotic trajectories between encounters are depicted as simple links.
Calculation of any given $\langle p_\lambda(T)\rangle$ requires constructing all
possible contributing diagrams. Most importantly, a diagrammatic rule has been found
for the value of any diagram. In the absence of tunnel barriers, the rule consists in three types of multiplicative factors: $M^{-1}$ for each link; $-M$ for each vertex; and $N_a$ for each trajectory beginning or ending in terminal $a$.

In Fig.~\ref{trajetorias}, we have two examples of these sets of trajectories. The blue trajectory has a 3-encounter, while the red trajectory has a 2-encounter (the red trajectory also has an encounter happening at the tunnel barrier; this will be discussed later). The pair consisting of the blue trajectory and its partner contributes $N_1N_2(-M)M^4=-N_1N_2/M^3$ to the semiclassical calculation of conductance, $\lambda=(1)$.

\subsection{Matrix integral for two terminals} 

As discussed in \cite{matrix}, for systems with broken time-reversal symmetry and two terminals, the sum over diagrams, with the correct diagrammatic rules, can be performed automatically by the following matrix integral
\begin{eqnarray}\label{integral0}\fl
\langle p_\lambda(T)\rangle &=& \lim_{N \rightarrow 0}\sum_{\vec{i}, \vec{o}}\frac{1}{\mathcal{Z}}\int \exp \left( -M\sum_{q = 1}^\infty \frac{1}{q}\mathrm{Tr} (Z^\dag Z)^q \right )\prod_{k = 1}^n Z_{o_ki_k}Z^*_{o_ki_{\pi(k)}}dZ,
\end{eqnarray}
where $Z$ is a complex square matrix of size $N$. The exponential function is responsible for producing all possible encounter structures inside the billiard, while the product of matrix elements of $Z$ and $Z^*$ represent the incoming and outgoing channels. Diagrammatic formulation of matrix integrals, based on the celebrated Wick theorem, is a well established topic \cite{zvon,diag2,diag3,morris,franc}.

Throughout this paper, we denote by $\mathcal{Z}$ a generic normalization constant, whose specific form changes depending on the integral being considered. In the equation above it is given by
\begin{equation}
\mathcal{Z}=\int e^{-M\mathrm{Tr} (Z^\dag Z)}dZ.
\end{equation}

Notice that the matrix size $N$ must be taken to zero after the integral has been performed. This is necessary because the diagrammatic expansion of the matrix integral involves closed loops that would physically correspond to periodic orbits, but such orbits are not allowed in the semiclassical approach to transport. Taking $N\to 0$ removes these spurious terms \cite{matrix}.   

In the presence of intact time-reversal symmetry, a corresponding matrix integral was introduced in \cite{trs}. There is then no distinction between $Z$ and $Z^*$, this matrix is real. This makes it a bit more complicated to differentiate between trajectories entering or leaving the billiard. In order to control the channel labels, one introduces  $R = YQZQY^\dag$ with $Q = \mathbb{I}_M \oplus 0_{N - M}$ a projector and $Y$ a $M \times M$ auxiliary matrix. We will be interest only in the simplest quantity, the average conductance $\sum_{io}\langle |t_{oi}|^2\rangle$. In that case this theory gives
\begin{equation}\label{tio}\fl
\langle t_{oi}t^*_{oi}\rangle= [W_{oi}W^*_{oi}]\lim_{N \rightarrow 0}\frac{1}{\mathcal{Z}}\int \exp \left( -\frac{M}{2}\sum_{q = 1}^\infty \frac{1}{q}\mathrm{Tr} (Z^T Z)^q \right )R_{ii}R_{oo}dZ,
\end{equation}
where $[W_{oi}W^*_{oi}]$ means we should extract the coefficient of $W_{oi}W^*_{oi}$ from whatever is the result of the calculation (matrix $W$ is defined as $W=YY^T$), and now
\begin{equation}
\mathcal{Z}=\int e^{-\frac{M}{2}\mathrm{Tr} (Z^T Z)}dZ.
\end{equation}

These semiclassical matrix integrals have been used not only to demonstrate equivalence of the semiclassical approximation to RMT calculations \cite{matrix,trs}, in a much more straightforward way than had previously been done, but also to go beyond random matrices in the treatment of energy-dependent correlations between elements of the $S$ matrix \cite{energy1, energy2}.

\subsection{Placing a tunnel barrier}\label{rmtintegral}

When a tunnel barrier is present, action correlations may be of a different nature, as trajectories that hit the barrier from the inside may fail to tunnel out and, instead, be reflected back into the billiard. When this happens a trajectory will be composed of two or more parts, corresponding to its excursions between hits on the barrier. Two trajectories may then differ in the order of these excursions, while still having the same action. An example is shown in Fig.~\ref{trajetorias}, consisting in the red trajectory and its black partner; this example contains three hits on the barrier.

The presence of the barrier modifies the diagrammatic rules. Let us say that the barrier is in terminal $a$. Given the barrier's reflectivity $\gamma$, the multiplicative factors become: $(M-\gamma N_a)^{-1}$ for each link; $-M+\gamma^qN_a$ for each $q$-encounter; $\gamma$ for each hit on the barrier; and $N_b-\gamma N_b\delta_{ab}$ for each trajectory beginning or ending in terminal $b$. In a system with three terminals, $M=N_1+N_2+N_3$.

Suppose transport is between terminals $1$ and $2$ and the barrier is in terminal $3$. Then, the pair with the blue trajectory in Fig.~\ref{trajetorias} contributes $(-M + N_3\gamma^3)$, due to the 3-encounter, $(M - N_3\gamma)^{-4}$ due to four links and $N_1N_2$ from entering and leaving the cavity. On the other hand, the pair with the red trajectory contributes $(-M + N_3\gamma^2)$ times $(M - N_3\gamma)^{-6}$ times $\gamma^3$ times $N_1N_2$. 

Bento  and  Novaes developed a matrix integral approach to transport with broken time-reversal symmetry, having two terminals and a tunnel barrier in terminal 1 \cite{pedro}. They introduced terms like $Z\left(\gamma Z^\dag Z\right)^k$ to represent a trajectory hitting the barrier $k$ times before leaving the billiard. As the number of hits can be arbitrary large, a sum over $k$ is needed, which produces a geometric series of $\gamma Z^\dag Z$. Their integral is
\begin{equation}\fl
\frac{1}{\mathcal{Z}}\int \exp \left( -\sum_{q = 1}^\infty \frac{M-N_1\gamma^q}{q}\mathrm{Tr} (Z^\dag Z)^q \right )\prod_{k = 1}^n Z_{o_ki_{\pi(k)}}^*\left(Z\frac{1}{1-\gamma Z^\dagger Z}\right)_{o_ki_k}dZ,
\end{equation}
with
\begin{equation}
\mathcal{Z}=\int e^{-(M-N_1\gamma)\mathrm{Tr} (Z^\dag Z)}dZ.
\end{equation}
Notice how the presence of the tunnel barrier affects the exponent and the matrix elements of $Z$ that correspond to terminal $1$. Of course, the same approach can also be used when the barrier is in terminal 2, provided one simply exchanges $N_1\leftrightarrow N_2$.

On the other hand, if we want to consider scattering from terminal 1 to terminal 2 in a cavity with three terminals, with a barrier in terminal $3$, the matrix integral should be modified only in the exponent:
\begin{equation}\label{integral3}\fl
\langle p_\lambda(T)\rangle = \lim_{N \rightarrow 0}\sum_{\vec{i}, \vec{o}}\frac{1}{\mathcal{Z}}\int \exp \left( -\sum_{q = 1}^\infty \frac{\left( M - N_3\gamma^q\right)}{q}\mathrm{Tr} (Z^\dag Z)^q \right )\prod_{k = 1}^n Z_{o_ki_{\pi(k)}}^*Z_{o_ki_k}dZ,
\end{equation}
with
\begin{equation}
\mathcal{Z}=\int e^{-(M-N_3\gamma)\mathrm{Tr} (Z^\dag Z)}dZ.
\end{equation}

\section{Results for broken Time-Reversal Symmetry}

In this section, we use the method discussed above to obtain expressions for average of conductance, variance of conductance, shot-noise power and third cumulant for different limits. We make use of a number concepts from combinatorics/representation theory. These are collected in the Appendix.

Results are confronted with calculations from diagrammatic RMT method in the limit of a large number of channels.We have checked that our results agree with numerical simulations done by sampling $10^5$ random scattering matrices in the presence of the tunnel barrier (using the approach from \cite{kumar2}). The curves obtained from the simulation (not shown) are indistinguishable from the theoretical ones.

\subsection{Transport between terminals 1 and 2, barrier in terminal 3}

In order to solve the matrix integral in Eq.(\ref{integral3}), we start by applying the singular value decomposition $Z = UDV$ with $U$ and $V$ belonging to the group $\mathcal{U}(N)$ of $N\times N$ unitary matrices, and $D$ is a real diagonal matrix. The jacobian of this change of variables is proportional \cite{jacobians} to $|\Delta(X)|^2$, in terms of the Vandermonde
\begin{equation}
    \Delta(X) = \prod_{i= 1}^{N-1}\prod_{j=i+1}^N (x_j - x_i) 
\end{equation}
of the matrix $X = D^2$ with eigenvalues $\{x_1,...,x_N\}$.

The normalization is
\begin{equation}
    \mathcal{Z}=\int e^{-(M-N_3\gamma)\mathrm{Tr}X}|\Delta(X)|^2dX.
\end{equation}
This is a particular case of the Selberg integral, Eq.(\ref{SelbExp}), which gives
\begin{equation}
    \mathcal{Z}=\frac{1}{(M-N_3\gamma)^{N^2}}\prod_{j=1}^N j!(j-1)!.
\end{equation}

The integrals over the unitary group,
\begin{equation}
\sum_{\vec{i}, \vec{o}}\int \prod_{k = 1}^n (UDV)^*_{o_ki_{\pi(k)}}(UDV)_{o_ki_k}dUdV.
\end{equation}
can be performed by making use of the machinery of the Weingarten functions and orthogonality relations of irreducible characters of the symmetric group, see \cite{collins,CS, matrix}. After summing over the incoming and outgoing channels, we arrive at
\begin{equation}
    \sum_{\mu\vdash n}\frac{[N_1]^{(1)}_{\mu}[N_2]^{(1)}_{\mu}}{\left([N]_{\mu}^{(1)}\right)^2}\chi_{\mu}(\pi)s_{\mu}(X),
\end{equation}
where $[N]^{(1)}_\lambda$ is a polynomial in $N$ given by Eq.~(\ref{stoN}) and $s_{\mu}(X)$ are the Schur polynomials .


The remaining integral over the diagonal matrices $X$ is 
\begin{equation}
I=\frac{1}{\mathcal{Z}}\int \det(1-X)^{M}\det(1-\gamma X)^{-N_3}s_\mu(X)|\Delta(X)|^2dX.
\end{equation}
This is best approached as a power series in $\gamma$. To that end, we employ the Cauchy identity, Eq.(\ref{cauchyschur}), to write $\det(1-\gamma X)^{N_3}$ as an infinite linear combination of Schur polynomials $s_\rho(X)$. This leads to the product $s_\rho(X)s_\mu(X)$. This can also be written as an infinite linear combination of Schur polynomials $s_\alpha(X)$, by means of the Littlewood-Richardson coefficients defined in Eq.~(\ref{lrcoef}). Putting all together and using the Selberg-Jack integral, Eq.~(\ref{jackselberg}), we arrive at 
\begin{equation}\fl
I=(M-N_2\gamma)^{N^2}\sum_\rho s_\rho(\gamma 1^{N_3})\sum_\alpha C^{(1)}_{\mu\rho\alpha}s_\alpha(1^N)[N]_\alpha^{(1)}\prod_{j=1}^N\frac{(M+N-j)!}{(\alpha_j+M+2N-j)!}.
\end{equation}

When $N\to 0$, the value of $I$ approaches
\begin{equation}
\sum_\rho s_\rho(\gamma 1^{N_3})\sum_\alpha C^{(1)}_{\mu\rho\alpha}\frac{d_\alpha}{|\alpha|!}\frac{([N]_\alpha^{(1)})^2}{[M]_\alpha^{(1)}},
\end{equation}
where we used the relation between $s_\alpha(1^N)$ and the polynomial $[N]_\alpha^{(1)}$ given in Eq.(\ref{stoN}) and $d_\alpha=\chi_\alpha(1^{|\alpha|})$ is the dimension of the irreducible representation of the permutation group labeled by $\alpha$. 

We need to know the value of $[N]_\alpha^{(1)}/[N]_\mu^{(1)}$ as $N\to 0$. This follows from the expression for $[N]_\alpha^{(1)}$ in terms of contents and the size of the Durfee square contained in $\alpha$, Eq.~(\ref{expanN}). Along with Eq.~(\ref{lrcoef}), this implies that the limit exists and is different from zero if, and only if, $D(\alpha) = D(\mu)$. It equals
\begin{equation}
\lim_{N\to 0}\frac{[N]_\alpha^{(1)}}{[N]_\lambda^{(1)}}=\frac{t_1(\alpha)}{t_1(\lambda)}\delta_{D(\alpha),D(\mu)},
\end{equation}
where $t_1(\lambda)$ is the product of all non-zero $1$-contents of $\lambda$.

We shall also use the fact that 
\begin{equation}
\sum_\rho C^{(1)}_{\mu\rho\alpha}s_\rho(\gamma 1^{N_3})=\gamma^{|\alpha|-|\mu|} s_{\alpha\backslash\mu}(1^{N_3}),
\end{equation}
in terms of a skew Schur function. This finally gives
\begin{equation}\fl
\langle p_\lambda(T)\rangle = \sum_{\mu\vdash n}[N_1]_{\mu}^{(1)}[N_2]^{(1)}_{\mu}\chi_{\mu}(\lambda)\sum_{\stackrel[D(\alpha)=D(\mu)]{}{\alpha \supset \mu}}  s_{\alpha\setminus \mu}(1^{N_3})\frac{d_\alpha\gamma^{|\alpha| -|\mu|}}{|\alpha|![M]^{(1)}_\alpha}\left( \frac{t_{(1)}(\alpha)}{t_{(1)}(\mu)}\right)^2.\label{pT}
\end{equation}
The above expression is perhaps not very easy on the eyes, but it is explicit, in the sense that many terms can be obtained exactly in a computer algebra system without difficulty. Below, we will use Eq. (\ref{pT}) to obtain general expression for electronic transport observables.


\subsection{Average and variance of conductance}

Let $g_{12}=p_{(1)}\left(T\right),$ be the conductance between terminals 1 and 2. From Eq. (\ref{pT}), we can show that 
\begin{equation}
\langle g_{12}\rangle = N_1N_2\sum_{\stackrel[D(\alpha)=1]{}{\alpha\supset (1)}} \gamma^{|\alpha|-1} s_{\alpha\setminus (1)}(1^{N_3})\frac{d_\alpha}{|\alpha|![M]_\alpha^{(1)}} \left(t_{(1)}(\alpha)\right)^2.
\end{equation}
The sum over $\alpha$ can be parameterized noticing that $\alpha$ must be a hook, i.e. of the form $\alpha = (m + 1 - k, 1^k)$, with $k = 0, \cdots, m$. In this case, $t_\alpha^{(1)} = (-1)^kk!(m-k)!$ and the skew Schur function $s_{(m + 1 - k, 1^k)\setminus (1)}(1^{N_3})$ can be replaced by a product of rising and falling factorials, Eq. (\ref{schurespecial}). At the end, we get a rather friendly expression:
\begin{equation}\label{t12ntrs}
\langle g_{12}\rangle= \frac{N_1N_2}{M}\sum_{m = 0}^\infty \frac{\gamma^{m}}{m + 1} \sum_{k=0}^m\frac{(N_3)^{m-k}(N_3)_{k}}{\left(M+1\right)^{m-k}\left(M-1\right)_{k}}.\label{gg}
\end{equation}
The conductance variance provides information about quantum chaos and the universality of electronic transport. 
We can obtain the average and variance of conductance as expansion in power series of $\gamma$ as follows 
\be
    \frac{\langle g_{12}\rangle}{N_1N_2} = \frac{1}{M} + \frac{N_3}{\left(M^2 - 1\right)}\gamma + \frac{N_3\left(MN_3 - 2\right)}{\left(M^2 - 1\right)\left(M^2 - 4\right)}\gamma^2 + \mathcal{O}\left(\gamma^3\right),\label{gG}\ee
    \be\fl
     \frac{\textbf{var}[g_{12}]}{N_1N_2} = \frac{\left(M -N_1\right) \left(M - N_2 \right)}{M^{2} \left(M^2 - 1\right)} + \frac{2N_3\left(M - 2N_1\right) \left(M - 2N_2 \right)}{M \left(M^2 - 1\right)\left(M^2 - 4
    \right)}\gamma+ \mathcal{O}(\gamma^2).\label{vargg}
\ee
Eqs. (\ref{gG}) and (\ref{vargg}) are valid for any value of channels $N_i=1,2,\cdots$, where $M=\sum_{i=1}^3N_i$. Besides, when $\gamma\rightarrow 0$ only the first term of Eq. (\ref{gG}) and (\ref{vargg}) survive, which is equivalent to a system connected ideally to terminals.

It is interesting to analyze some experimental limits. In the first one we take $N_1=N_2=N_3 = N_0$. Then, Eqs. (\ref{gG}) and (\ref{vargg}) simplify to
\be\fl
    \langle g_{12}\rangle = \frac{N_0}{3} + \frac{N_0^{3}\gamma}{9 N_0^2 -1}
    + \frac{N_0^{3} \left(3 N_0^{2}-2\right) \gamma^2}{\left(9 N_0^2 -1\right) \left(9 N_0^2 -4\right) }
    +\mathcal{O}(\gamma^3),\label{g1}\ee
    \be\fl
     \textbf{var}[g_{12}] = \frac{4}{9}\frac{N_0^{2}}{9N_0^2 - 1} + \frac{2}{3}\frac{ N_0^{4} \gamma}{\left(9 N_0^2 -1\right) \left(9 N_0^2 -4\right)}+ \frac{1}{3}\frac{N_0^{4} \gamma^2}{  \left(9 N_0^2 -1\right)^{2} \left(9 N_0^2 - 4\right)}+ \mathcal{O}(\gamma^3).\label{varg1}
\ee
Taking the limit of a large number of channels ($N_0\gg 1$), we get
\begin{eqnarray}
    \langle g_{12}\rangle &=& \frac{N_0}{3} + \frac{N_0}{9}\gamma
    + \frac{N_0 }{27}\gamma^2
    +\mathcal{O}(\gamma^3),\label{g11}\\
     \textbf{var}[g_{12}] &=& \frac{4}{81} + \frac{2}{243}\gamma+ \frac{1}{243}\gamma^2+ \mathcal{O}(\gamma^3).\label{varg11}
\end{eqnarray}
Eqs. (\ref{g11}) and (\ref{varg11}) are in agreement with results calculated from RMT, which will be discussed in Section \ref{rmt}. 

As the second limit of Eqs. (\ref{gG}) and (\ref{vargg}), we keep the number of channels in the terminals 1 and 2 in the extreme quantum regime, $N_1=N_2=1$, while $N_3$ is arbitrary. This limit is currently inaccessible to known literature results with multiterminals and tunnel barriers. We obtain  
\be\fl
     \langle g_{12}\rangle = \frac{1}{N_3+2} + \frac{N_3}{\left(N_3 +1\right) \left(N_3 +3\right)}\gamma + \frac{\left(N_3^{2}+2 N_3 -2\right) }{\left(N_3 +1\right)\left(N_3 +3\right) \left(N_3 +4\right) }\gamma^{2} +  \mathcal{O}(\gamma^3),\label{g2}\ee 
     \be\fl
     \textbf{var}[g_{12}]  = \frac{N_3 +1}{\left(N_3 +2\right)^{2} \left(N_3 +3\right)}+\frac{2 N_3^{2}}{\left(N_3 +1\right) \left(N_3 +2\right) \left(N_3 +3\right) \left(N_3 +4\right) }\gamma + \mathcal{O}(\gamma^2).\label{varg2}
\ee
When $N_3\gg1$, the equations above simplify to
\begin{eqnarray}
     \langle g_{12}\rangle &=& \frac{1}{N_3}\sum_{k = 0}^\infty \gamma^k= \frac{1}{N_3\left(1 - \gamma\right)}\\ 
     \textbf{var}[g_{12}]  &=&\frac{1}{N_3^2}\sum_{k = 0}^\infty \left( k +1 \right)\gamma^k= \frac{1}{N_3^2\left(1 - \gamma\right)^2}. 
\end{eqnarray}
Note that the equations above are valid for the range $0\leq\gamma\leq1$.
When $\gamma\rightarrow1$ and $N_3 \rightarrow \infty$, with the product $N_3(1-\gamma)$ kept finite, this is known as the opaque limit \cite{whitney}. 

Moreover,  we can keep terminal 3 in the extreme quantum regime, $N_3=1$, while $N_1=N_2=N_0$. In this case, we have
\be
    \langle g_{12}\rangle = \frac{N_0^2}{2N_0+1} + \frac{1}{4}\frac{N_0}{N_0 + 1}\gamma + \frac{1}{4}\frac{N_0 }{ \left(N_0 +1\right) \left(2 N_0 + 3 \right)}\gamma^{2}+\mathcal{O}(\gamma^3)\label{g3}\ee
    \be
     \textbf{var}[g_{12}]  = \frac{1}{4}\frac{N_0 \left(N_0 +1\right)}{\left(2 N_0 +1\right)^{2}} +\frac{1}{2}\frac{ N_0}{  \left(N_0 +1\right) \left(2 N_0 +3\right) \left(4 N_0^2 -1\right)}\gamma + \mathcal{O}(\gamma^2).\label{varg3}
\ee
When $N_0\gg1$, these simplify to
\begin{eqnarray}
     \langle g_{12}\rangle &=&\frac{N_0}{2}-\frac{1}{4} + \frac{1}{4}\gamma + \frac{1 }{8N_0    }\gamma^{2} +  \mathcal{O}(\gamma^3),\\
     \textbf{var}[g_{12}]  &=& \frac{1}{16} - \frac{1}{16N_0}\gamma^2 +\frac{1}{16N_0}\gamma^3 + \mathcal{O}(\gamma^4).
\end{eqnarray}

\begin{figure*}
    \centering
    \includegraphics[scale = 0.4]{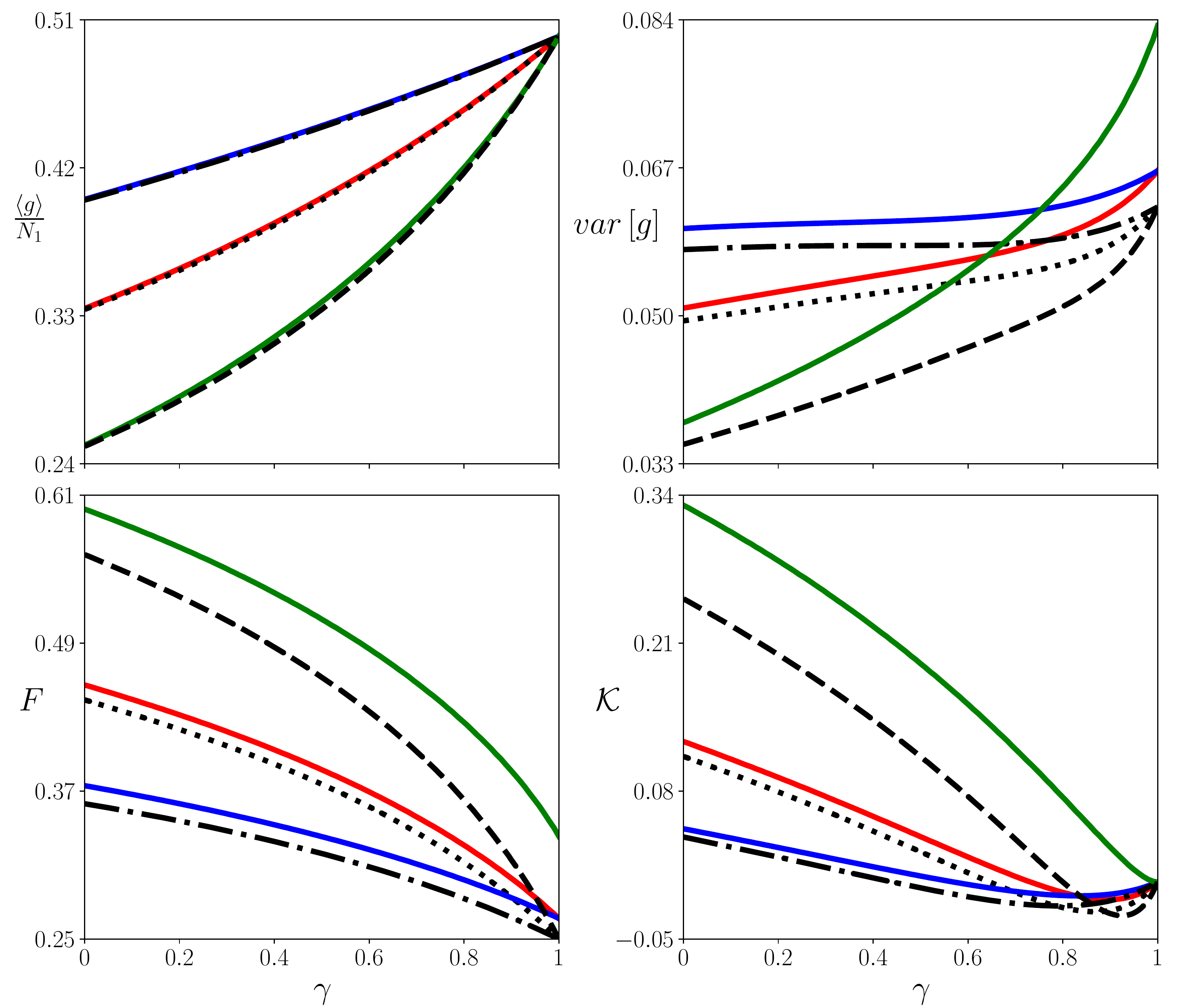}
    \caption{(color online)Average conductance (left top panel), conductance variance  (right top panel), Fano factor (left down panel) and third cumulant (right down panel) as functions of $\gamma$, for broken time-reversal symmetry. Solid lines are our results for $N_0 = 2$, while broken lines are asymptotic $N_0 \to\infty$ RMT results. Red/dotted lines are for $N_1 = N_2 = N_3 = N_0$; blue/dash-dotted lines are for $N_1 = N_2 = N_0$ and $N_3 = 1$; green/dashed lines are for $N_1= N_2 = 1$ and $N_3 = N_0$. We sum the series until convergence. Numerical simulations are indistinguishable from theoretical results.}
    \label{moments_cases}
\end{figure*}

We plot in Fig. \ref{moments_cases} and Fig. \ref{fig:my_label} the average and variance of conductance as functions of $\gamma$ in several different regimes. Channel numbers are always small in Fig. \ref{moments_cases}, while they are always equal in Fig. \ref{fig:my_label}. Also shown are RMT results (black lines), which are valid for large $M$, see Section \ref{rmt}. The average conductance is always close to its asymptotic value, even for the smallest channel numbers, but the same is not true for higher transport moments.

\subsection{Higher order moments}

Eq.~(\ref{pT}) also enables us to compute higher transport moments. However, in contrast to the conductance, these moments are more cumbersome. Specifically, we are interested in the average shot noise power $\langle p\rangle$, which carries information about time-dependent fluctuations in the electrical current due to the discreteness of the electrical charge, and in the third cumulant, 
\begin{equation}
k =  p_{(1)}\left(T\right)  - 3 p_{(2)}\left(T\right)  + 2p_{(3)}\left(T\right) .\label{k}
\end{equation}
which has also attracted some recent interest, including an experimental observation in tunnel junctions \cite{Reulet,bar1,anderson3}.

Moreover, we may consider related quantities as the Fano factor, which is the ratio between the averages of shot-noise power and the conductance, $$F = \frac{\langle p\rangle}{\langle g_{12}\rangle},$$ and ratio between the averages of third cumulant and shot-noise power $$\mathcal{K} = \frac{\langle k \rangle}{\langle p\rangle}.$$ 
As done above, we present the first few terms of $F$ and $\mathcal{K}$ for symmetric channel numbers, $N_1=N_2=N_3=N_0$:
\begin{eqnarray}
    F &=& \frac{4 N_0^{2}}{9 N_0^2 -1}-\frac{9 N_0^{4} \left(9 N_0^{2}-5\right) }{\left(9 N_0^2 -1\right)^{2} \left(9 N_0^2 -4\right)}\gamma+ \mathcal{O}(\gamma^2), \label{F1}\\
    \mathcal{K} &=& \frac{N_0^{2}}{9 N_0^2 - 4}-\frac{1}{2}\frac{24 N_0^{4}-19 N_0^{2}+4}{\left(9N_0^2 - 4\right)^{2}} \gamma +  \mathcal{O}(\gamma^2).\label{K1}
\end{eqnarray}
When $N_0\gg1$, these simplify as
\begin{eqnarray}
    F &=& \frac{4}{9}-\frac{1}{9} \gamma -\frac{1}{27} \gamma^{2}+\frac{2}{81} \gamma^{3} + \mathcal{O}(\gamma^4), \label{F11}\\
    \mathcal{K} &=& \frac{1}{9}-\frac{4}{27} \gamma -\frac{1}{27} \gamma^{2}-\frac{11}{972} \gamma^{3} + \mathcal{O}(\gamma^4),\label{K11}
\end{eqnarray}
in agreement with RMT results from Section \ref{rmt}.
As the second limit, we take $N_1=N_2=1$ to obtain
\begin{eqnarray}
    F &=&  \frac{N_3 +1}{N_3 +3}-\frac{2 N_3 \left(N_3 +2\right)^{2} }{\left(N_3 +1\right)\left(N_3 +3\right)^{2} \left(N_3 +4\right) }\gamma +  \mathcal{O}(\gamma^2), \\ 
    \mathcal{K} &=&  \frac{N_3}{N_3 +4}-\frac{4 N_3 \left(N_3 +2\right)^{3}}{\left(N_3 +1\right)^{2} \left(N_3 +4\right)^{2} \left(N_3 +5\right)}\gamma
    + \mathcal{O}(\gamma^2).
\end{eqnarray}
When $N_3\gg 1$, these simplify as
\begin{eqnarray}
    F &=& 1- \frac{2}{N_3}\sum_{k = 0} \gamma^k =  1-\frac{2}{N_3 \left(1-\gamma \right)}, \\ 
    \mathcal{K} &=& 1- \frac{4}{N_3}\sum_{k = 0} \gamma^k=  1-\frac{4   }{N_3 \left(1-\gamma \right)}. 
\end{eqnarray}
Besides, when $N_3=1$ and $N_1=N_2=N_0$, we have
\be
    F =  \frac{1}{4}\frac{N_0 +1}{ N_0}-\frac{1}{16}\frac{\left(2 N_0^{2}+3 N_0 -3\right) \left(2 N_0 + 1\right)^{2} }{  N_0^{2} \left(N_0 +1\right) \left(2 N_0 - 1\right)\left(2 N_0 + 3\right)}\gamma + \mathcal{O}(\gamma^2), \ee
    \be\fl
    \mathcal{K} =\frac{1}{\left(2 N_0 - 1\right) \left(2 N_0 + 3\right)}-\frac{1}{2}\frac{\left(2 N_0 + 1\right) \left(4 N_0^{4}+8 N_0^{3}+N_0^{2}-3 N_0 +2\right)}{ \left(N_0 +1\right)^{2}\left(2 N_0 - 1\right)^{2} \left(2 N_0 + 3 \right)^{2} } \gamma + \mathcal{O}(\gamma^2).
\ee
When $N_0\gg 1$, these simplify as
\begin{eqnarray}
    F &=&  \frac{1}{4} -\frac{1}{8N_0}\gamma -\frac{1}{8N_0} \gamma^2 -\frac{3}{16N_0^2} \gamma^3 + \mathcal{O}(\gamma^4), \\ 
    \mathcal{K} &=&  \frac{1}{4N_0^2}-\frac{1}{4N_0}\gamma -\frac{1}{8N_0^2}\gamma^2 + \frac{1}{4N_0}\gamma^3
    + \mathcal{O}(\gamma^4).
\end{eqnarray}

\begin{figure*}
    \centering
    \includegraphics[scale = 0.4]{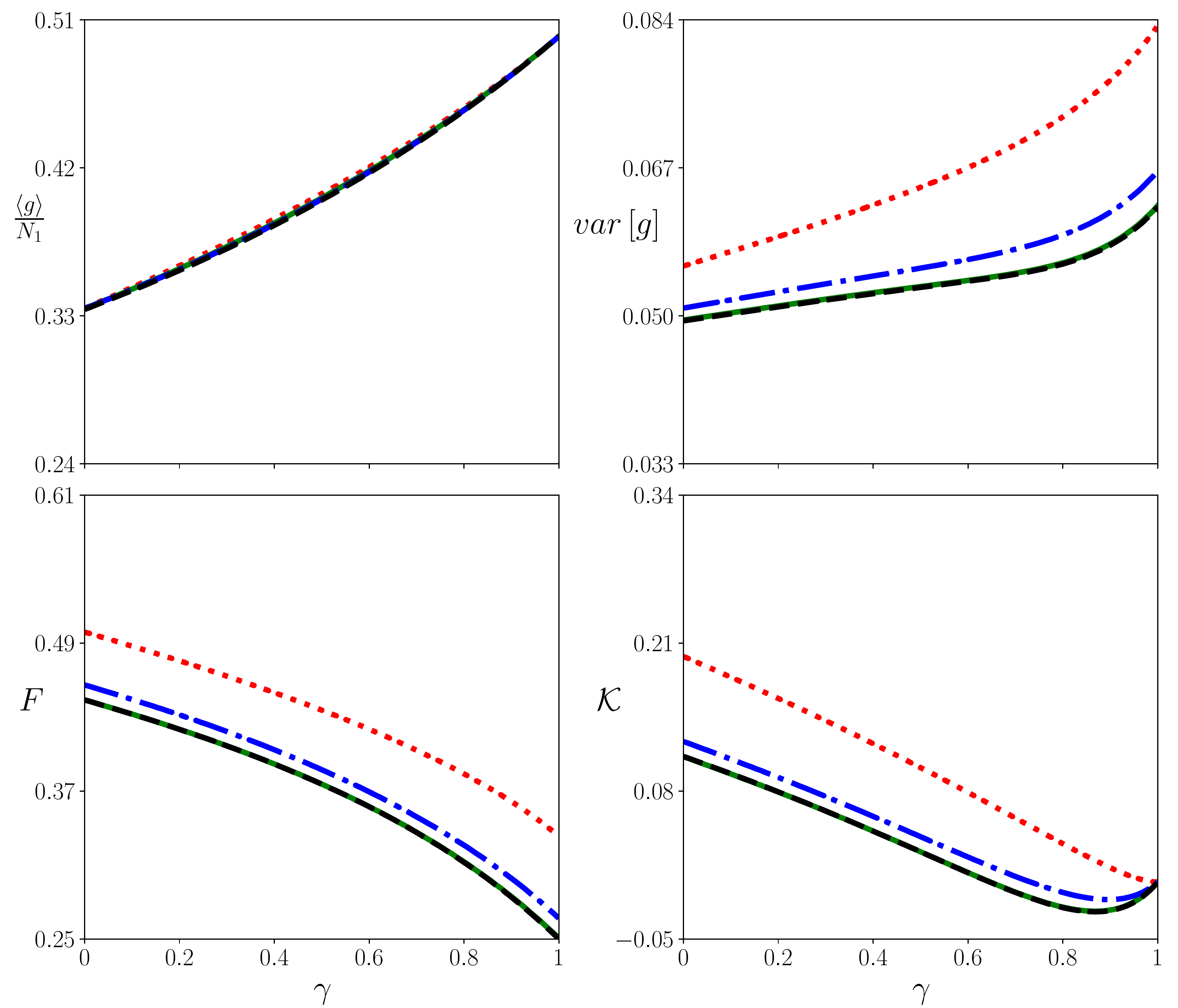}
    \caption{Average of conductance (left top panel), conductance variance (right top panel), Fano factor (left down panel) and third cumulant (right down panel) as functions of $\gamma$ for symmetric terminals $N_1 = N_2 = N_3 = N_0$ and broken time-reversal symmetry. In all panels, we have $N_0 = 1$ for the red dotted line,  $N_0 = 2$ for the blue dash-dotted line and $N_0\to infty$ for the black RMT prediction. The green solid line corresponds to $N_0 = 5$ in the top left panel, $N_0 = 12$ in the top right panel, $N_0 = 8$ in the left down panel  and $N_0 = 10$ in the right down panel. We sum the series until convergence. Numerical simulations are indistinguishable from theoretical results.}
    \label{fig:my_label}
\end{figure*}

We plot in Fig. \ref{moments_cases} and Fig. \ref{fig:my_label} the average and variance of conductance as functions of $\gamma$ in several different regimes. Channel numbers are always small in Fig. \ref{moments_cases}, while they are always equal in Fig. \ref{fig:my_label}. Also shown are RMT results (black lines), which are valid for large $M$, see Section \ref{rmt}. The average conductance is always close to its asymptotic value, even for the smallest channel numbers. Higher moments show more noticeable differences, i.e. quantum corrections are more important. 

Fig. \ref{moments_cases} and Fig. \ref{fig:my_label} show $F$ and $\mathcal{K}$ as functions of $\gamma$. For higher moments the differences between our results (extreme quantum regime) and the RMT (asymptotic) results are more noticeable, specially for very small channel numbers. The discrepancy does not seem to depend much on the value $\gamma$.

The problem of convergence of the series is addressed in Fig.4. We choose the worse value for convergence, $\gamma=1$, and the symmetric situation $N_1=N_2=N_3=N_0$. The corresponding series for average conductance and third cumulant, (\ref{t12ntrs}) and (\ref{K1}), are computed up to order $n_{max}$ and the result is shown as a function of $n_{max}$. Average conductance converges with about $8$ terms, for both $N_0=2$ and $N_0=10$, while $\mathcal{K}$ requires more terms, around $25$.

\begin{figure*}
    \centering
    \includegraphics[scale = 0.4]{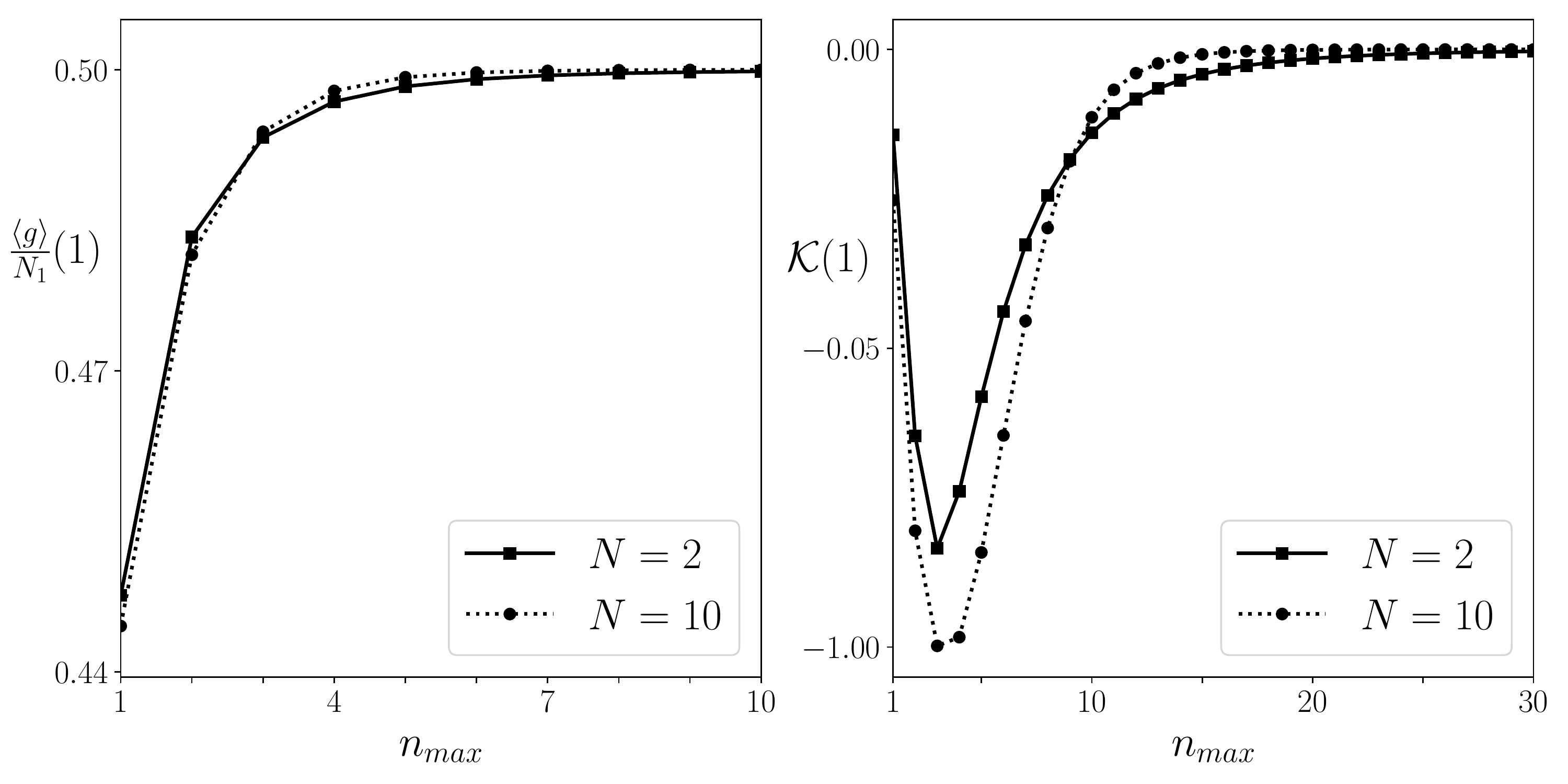}
    \caption{Value of $\langle g\rangle$ and $\mathcal{K}$ at $\gamma=1$, as functions of the number of terms included in the corresponding sums over powers of $\gamma$. Here we take $N_1=N_2=N_3=N_0$. We see convergence in both cases and for two different values of $N_0$.}
    \label{convergence}
\end{figure*}

\subsection{Random Matrix Theory}\label{rmt}

In order to confirm the validity of the results in the previous Section, we developed a diagrammatic calculation in the framework of RMT \cite{brouwer}, as developed by Brouwer and Beenakker. However, the results are limited to $M\gg1$. We are interested in broken time-reversal symmetry. From the diagrams of Ref. \cite{brouwer}, we were able to obtain the average and variance of conductance with three terminals and one barrier as shown in Fig. (\ref{trajetorias}). 

Taking $N_1=N_2=N_3 = N_0$, we have
\begin{eqnarray}
     \langle g_{12} \rangle&=& \frac{N_0}{3 - \gamma },\label{Sg}
 \\
    \textbf{var}[g_{12}] &=& \frac{4 \gamma^{4}-18 \gamma^{3}+48 \gamma^{2}-66 \gamma +36}{\left(\gamma - 3 \right)^{6}}. \label{Svarg}
\end{eqnarray}
Using the diagrams from Ref. \cite{anderson}, we obtain that the Fano factor is
\begin{eqnarray}
    F = \frac{\gamma^{3}-6 \gamma^{2}+15 \gamma -12}{\left(\gamma - 3 \right)^{3}},\label{FF}
\end{eqnarray}
which is in agreement with Ref. \cite{whitney2}.

Finally, to calculate the average of third cumulant we had to develop the diagrams for $ \langle\textbf{Tr}\left[\left(t_{12}t_{12}^\dagger\right)^3\right]\rangle$, which were not previously available in the literature. From these new diagrams (not shown in this work), we obtain that
\begin{eqnarray}
    \mathcal{K} = 
    \frac{\left(\gamma - 1 \right) \left(\gamma^{5}-8 \gamma^{4}+36 \gamma^{3}-78 \gamma^{2}+93 \gamma -36\right)}{\left(\gamma - 3 \right)^{3} \left(\gamma^{3}-6 \gamma^{2}+15 \gamma -12\right)}.\label{KK}
\end{eqnarray}
Expanding in power series of $\gamma$, we recover Eqs. (\ref{g11}), (\ref{varg11}), (\ref{F11}) and (\ref{K11}), respectively. Hence, we confirm the agreement between the both methods, in the limit $M\gg1$, for the first few orders in $\gamma$. However, for low values of $M$ (quantum limit) the semiclassical results can be markedly different from the asymptotic RMT results.

\section{Results for Intact Time-Reversal Symmetry}

Inclusion of a tunnel barrier in the semiclassical matrix model for systems with time-reversal symmetry is currently a challenge. But we can make progress in the particular case of conductance.

\subsection{Average conductance and quantum interference correction }

The presence of the tunnel barrier only changes the exponent of the matrix integral in Eq.(\ref{tio}), so we have to solve
\begin{equation}
 \lim_{N \rightarrow 0} \frac{1}{\mathcal{Z}}\int \exp\left(-\frac{1}{2} \sum_{q = 1}^\infty \frac{(M - \gamma^q N_3)}{q}\mathrm{Tr}(Z^TZ)^q\right) R_{ii}R_{oo} dZ.
\end{equation}

We introduce the singular value decomposition, $Z = ODP$, but now $O$ and $P$ are real orthogonal matrices. Orthogonality of matrix elements in the orthogonal group,
\begin{equation}
\int O_{ab}O_{cd}dO = \frac{\delta_{ac}\delta_{bd}}{N}
\end{equation} gives
\begin{equation}
\sum_{cd} D_{cc}D_{dd}\int O_{a_1c}O_{a_2d}dO \int P_{c b_1}P_{d b_2} dP = \frac{\mathrm{Tr}X}{N^2}.
\end{equation}
After this is taken care of, it remains to compute the eigenvalue integral
\begin{equation}\label{inttrs}
\frac{1}{\mathcal{Z}}\int \det{(1-X)}^{\frac{M}{2}}\det{(1 -\gamma X)}^{-\frac{N_3}{2}} \mathrm{Tr}X \frac{\Delta(X)}{\sqrt{\det(X)}}dX.
\end{equation}

The procedure now is parallel to what was done in Section 3.1. We use the Cauchy identity in terms of zonal polynomials, Eq.(\ref{cauchyzonal}), and the appropriate Littlewood-Richardson coefficients, Eq.(\ref{lrcoefzonal}). Knowing that $\mathrm{Tr}X=Z_1(X)$, we end up with a Selberg-Jack integral. 

The $N\to 0$ limit involves the ratio of two zonal polynomials, which can be expressed in terms of contents according to Eq.(\ref{expanN}). The limit tells us that the 2-Durfee square of the partition $\mu$ must have width $1$, therefore $\mu$ must be a double-hook. Together with Eq.~(\ref{lrcoefzonal}), we have that $D_2(\mu) = D_2(\rho)$, which implies that $\rho$ is also a double-hook, and can be parameterized as $(k_1, k_2, 1^{m - k_1 - k_2})\vdash m$. Using this parameterization it is possible to obtain closed expressions for $C_{\rho (1)\mu}^{(1)}$ \cite{stanley}. Summing over the incoming and outgoing channels we get
\begin{equation}
\langle g_{12}\rangle = N_1 N_2 \sum_{\stackrel[|\mu| = |\rho| + 1]{}{\mu,\rho}} \gamma^{|\rho|} C^{(2)}_{\rho (1)\mu} \frac{\left( t^{(2)}_\mu\right)^2}{j_\rho} \frac{Z_\rho(1^{N_3})}{\left[ M +1\right]_\mu^{(2)}}.\label{gTRS}
\end{equation}

The first few terms are given by
\begin{equation}\fl
   \langle g_{12}\rangle = \frac{N_1 N_2}{M +1} +\frac{N_1N_2N_3}{M \left(M +3\right) }\gamma + \frac{N_{1} N_{2} N_{3} \left(M N_{3}+M +N_{3}-3\right)}{ M\left(M -1\right) \left(M +3\right)\left(M +5\right)}\gamma^2+ \mathcal{O}\left(\gamma^3\right).\label{gT}
\end{equation}
Besides, from Eqs. (\ref{g1}) and (\ref{gT}) we are able to calculate the quantum interference correction of conductance, defined as the difference between the averages of conductance with intact TRS and broken TRS. We have
\begin{equation} 
   \langle \delta g\rangle = -\frac{N_{1} N_{2} }{M\left(M +1\right) }-\frac{N_{1} N_{2} N_{3} \left(3 M +1\right)}{  M \left(M +3\right)\left(M^2 -1\right)}\gamma + \mathcal{O}\left(\gamma^2\right).\label{dgT}
\end{equation}

When $N_1=N_2=N_3 = N_0$, Eqs. (\ref{gT}) and (\ref{dgT}) simplify to
\be\fl
    \langle g_{12}\rangle = \frac{N_0^{2}}{3 N_0 +1} +\frac{N_0^{2} }{3 \left(3 N_0 +3\right)}\gamma
     + \frac{N_0^{2} \left(3 N_0^{2}+4 N_0 -3\right) }{3 \left(3 N_0 -1\right) \left(3 N_0 +3\right)\left(3 N_0 +5\right) }\gamma^{2} +\mathcal{O}(\gamma^3),\label{gT1}\ee
     \be
     \langle \delta g\rangle = -\frac{N_0}{3\left( 3N_0 +1\right)}-\frac{N_0^{2} \left(9 N_0 + 1\right)}{3  \left(3 N_0 +3\right) \left(9 N_0^2 -1\right)}  \gamma +  \mathcal{O}(\gamma^2).\label{dgT1}
\ee
Considering a large number of channels ($N_0\gg 1$) leads to
    \be
     \langle \delta g \rangle = -\frac{1}{9} -\frac{1}{9} \gamma  -\frac{1}{27} \gamma^{2} -\frac{1}{243} \gamma^{3} + \mathcal{O}(\gamma^4,N_0^{-1}).\label{dgT11}
\ee
If $N_1=N_2=1$, we get
\be\fl
     \langle \delta g \rangle  = -\frac{1}{\left(N_3 +2\right)\left(N_3 +3\right) }   -\frac{N_3 \left(3N_3 +7\right) }{ \left(N_3 +1\right) \left(N_3 +2\right)\left(N_3 +3\right) \left(N_3 +5\right)}\gamma
  + \mathcal{O}(\gamma^2).\label{dgT2} 
\ee
When $N_3\gg 1$, this simplifies to
\be\fl
     \langle \delta g \rangle  =- \frac{1}{N_3^2}\sum_{k = 0}^\infty \left( 2k +1 \right)\gamma^k= -\frac{1 + \gamma}{N_3^2\left(1 - \gamma\right)^2}.
\ee
When $N_3=1$, while $N_1=N_2=N_0$, we have
    \be\fl
     \langle \delta g \rangle  = -\frac{N_0^{2}}{2 \left(N_0 +1\right) \left(2 N_0 +1\right)} -\frac{ N_0 \left(3 N_0 + 2\right) }{2\left( N_0 + 1\right) \left(2 N_0 +1\right)\left(2 N_0 +4\right) }\gamma + \mathcal{O}(\gamma^2).\label{dgT3}
\ee
When $N_0\gg1$, this becomes
\be
     \langle \delta g \rangle  =-\frac{1}{4}-\frac{3 }{8 N_0}\gamma+\frac{1}{8 N_0}\gamma^{2}+\frac{3 }{8 N_0^{2}}\gamma^{3} +  \mathcal{O}\left(\gamma^4, N_0^{-1}\right) 
\ee

To confirm the results above, we developed a diagrammatic calculation from RMT \cite{brouwer}, as discussed in \ref{rmt}, to obtain the quantum interference correction of conductance in the regime $M\gg1$, known as weak localization. For $N_1=N_2=N_3=N_0$, we obtain that
\begin{eqnarray}
\langle \delta g \rangle = -\frac{\gamma^2-3}{\left(\gamma-3\right)^3}, \label{RMTDG}
\end{eqnarray}
in agreement with Eq. (\ref{dgT11}).

\begin{figure*}
    \centering
    \includegraphics[scale = 0.35]{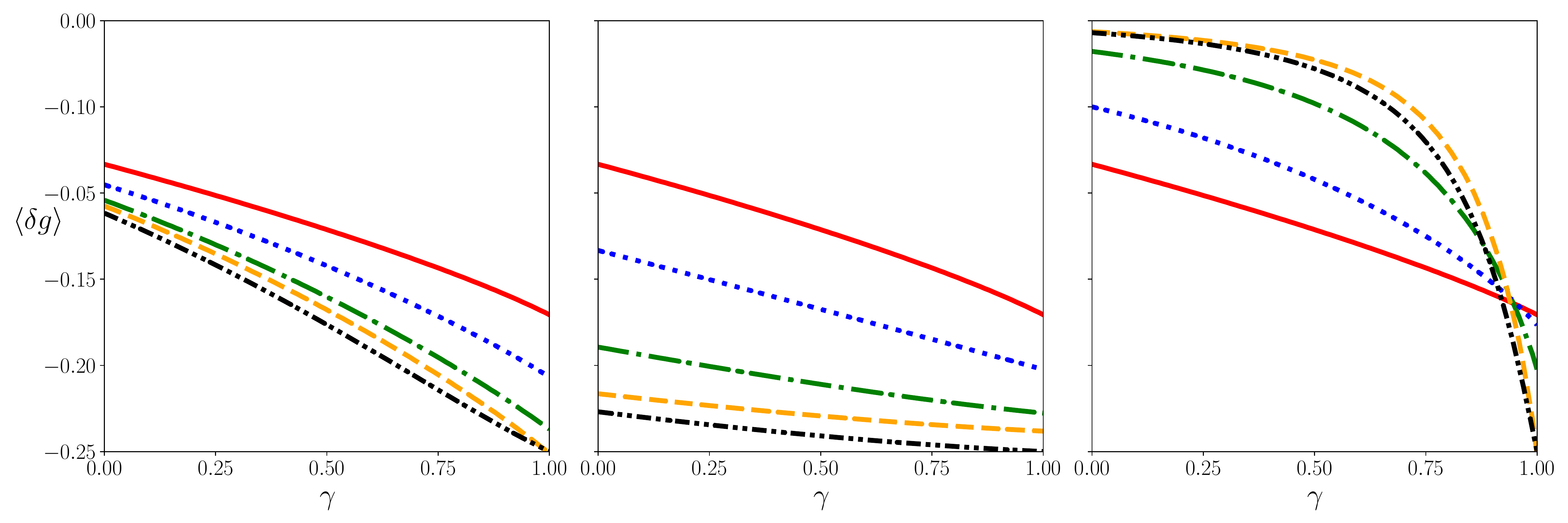}
    \caption{ Quantum interference correction of conductance as a function of the reflection rate $\gamma$. In the left panel, $N_1=N_2=N_3 = N_0$; in the central panel, $N_1=N_2 = N_0$ and $N_3 = 1$; and in the right panel $N_1 = N_2 = 1$ and $N_3 = N_0$. In all panels, the red solid line represents $N_0=1$; the dotted blue line represents $N_0=2$; the dash-dotted green line represents $N_0=5$; the dashed orange line represents $N_0=10$; and the dashed double dotted black line represents the RMT. Here $\delta g$ is expanded up to $\mathcal{O}\left(\gamma^5\right)$.}
    \label{WL}
\end{figure*}

Finally, Fig. (\ref{WL}) shows the quantum interference correction of conductance as a function of the reflection rate $\gamma$ for different number of channels. We can see that in all case the quantum interference correction decreases when $\gamma \to 1$. Besides, with increasing $N_0$ we have a good agreement between our results and RMT.


\subsection{Average conductance between terminals 1 and 3}

If we consider the tunnel barrier in the third terminal considered as the entering terminal, then it is required to introduce a geometric series inside $R$, by defining
\begin{equation}
    R^\prime = YQ\left( \frac{1}{1 - \gamma Z^\dag Z}Z\right) QY^\dag.
\end{equation}
The semiclassical matrix integral required for conductance then becomes
\begin{equation}
\lim_{N \rightarrow 0}\frac{1- \gamma}{\mathcal{Z}}\int \exp\left(-\frac{1}{2} \sum_{q = 1}^\infty \frac{(M - \gamma^q N_3)}{q}\mathrm{Tr}(Z^TZ)^q\right) R_{ii}^\prime R_{oo} dZ.
\end{equation}

The integrals over the orthogonal group coming from the singular value decomposition lead to
\begin{equation}
    \int  R_{ii}^\prime R_{oo} dOdP = W_{oi}W^*_{oi}  \sum_{k=0}^\infty \frac{\gamma^k }{N^2} \mathrm{Tr}\left( X^{k + 1}\right).
\end{equation}

The trace of $X^{k + 1}$ can be expressed as a sum over zonal polynomials
\begin{equation}
\mathrm{Tr}\left (X^{k+1}\right )= \frac{1}{(2k + 1)!} \sum_{\lambda \vdash k + 1} d_{2\lambda} t^{(2)}_\lambda Z_\lambda(X).
\end{equation}

The remaining integral over $X$ is
\begin{equation}
\frac{1}{\mathcal{Z}}\int \det{(1-X)}^{\frac{M}{2}}\det{(1 -\gamma X)}^{-\frac{N_3}{2}} Z_\lambda(X) \frac{\Delta(X)}{\sqrt{\det (X)}}dX.
\end{equation}
This integral is similar to Eq.~(\ref{inttrs}). Proceeding in the same way, we get 
\begin{equation}\fl
\langle g_{13}\rangle = (1- \gamma)N_1N_3\sum_{k=0}^\infty \sum_{\mu, \lambda}      \left( \sum_\rho \frac{\gamma^{|\rho|+ k}}{j_\rho}C^{(2)}_{\rho \lambda\mu} Z_\rho( 1^{N_3}) \right)  d_{2\lambda} \frac{t_\lambda^{(2)}}{\left( 2k + 1 \right)!}  \frac{\left( t^{(2)}_{\mu}\right)^2}{\left[ M +1\right]_\mu^{(2)}},\label{g13}
\end{equation}
where $\mu$ and $\lambda$ are double hooks, $\rho, \lambda \subset \mu$ and $|\mu| = |\rho| + |\lambda|$. The first few terms in the expansion are
\begin{equation}\fl
\frac{\langle g_{13}\rangle}{N_1N_3}= \frac{1}{M + 1} - \frac{\left( M - N_3 + 1 \right)}{M\left( M+3\right)}\gamma -\frac{ \left(N_{3} M +M +N_{3}-3\right) \left(M -N_{3}+1\right)}{M\left(M -1\right) \left(M +3\right)\left(M +5\right)  }\gamma^2+ \mathcal{O}\left(\gamma^3\right).
\end{equation}

\section{Dephasing regime \label{dephasingsec}}

As an application of our results, we are able to analyze the dephasing regime. A simple dephasing model was developed by Buttiker \cite{buttiker1}, who assumed that the current in terminals 1 and 2 hold the relation $I_1=-I_2=I$, while the current in terminal 3 is kept null $I_3=0$. That means that before escaping through terminals 1 or 2, the particle can escape through terminal 3 and come back, spending a long time inside the billiard, causing the dephasing. In other words, the dwell time can be much longer than the dephasing time, $\tau_d\gg\tau_\phi$. Applying these conditions to the Landauer-Buttiker approach \cite{landauer, buttiker}, the conductance between terminals 1 and 2 in the dephasing regime can be written as
\begin{equation}\label{conductancedf}
g = g_{21} + \frac{g_{23}g_{31}}{g_{32} + g_{31}},
\end{equation}
where $g_{ij}$ is the conductance from terminal $j$ to terminal  $i$. 
To obtain the average conductance in this regime, we need to take the average of Eq.~(\ref{conductancedf}). 
As discussed in \cite{baranger}, we consider the approximation of replacing each $g_{ij}$ with its mean $\left\langle g \right\rangle$,
\begin{equation}\label{Gdp}
\langle g \rangle = \langle g_{21} \rangle + \frac{\langle g_{23}\rangle \langle g_{31}\rangle}{\langle g_{32} \rangle + \langle g_{31}\rangle},
\end{equation}
which is true when $M\gg 1$.

\subsection{Broken Time-Reversal Symmetry}

Let's start by analyzing of the dephasing regime in the absence of TRS. The average conductance  $\langle g_{12}\rangle$ was obtained in Section 3.2, Eq. (\ref{gg}). On the other hand, from the results in \cite{pedro}  we have
 \begin{equation}\fl
 \langle g_{i3}\rangle= (1-\gamma)\frac{N_iN_3}{M}\sum_{m = 0}^\infty \frac{\gamma^{m}}{m + 1} \sum_{k=0}^m\frac{(N_3 + 1)^{m-k}(N_3- 1)_{k}}{\left(M+1\right)^{m-k}\left(M-1\right)_{k}}, \quad i = 1,2,\label{ggi3}
\end{equation}
which satisfies the relation $\langle g_{i3}\rangle = \langle g_{3i}\rangle$. Substituting Eqs. (\ref{gg}) and (\ref{ggi3}) in Eq. (\ref{Gdp}), we have the surprisingly simple result, obtained with the use of a computer:
\begin{equation}
    \langle g \rangle = \frac{N_1N_2}{N_1 + N_2}+ \mathcal{O}(\gamma^{21}).\label{gtrsp}
\end{equation}
Even though this looks very familiar, it is not obvious at all that it should hold for arbitrary $\gamma$. We could not prove this expression exactly, but in practice it seems that all orders in $\gamma$ vanish exactly. That means that the result is actually independent of $\gamma$. As $M=N_1+N_2+N_3\gg 1$, we can take that $N_3\gg1$, while $N_1=N_2=1$, which means that Eq. (\ref{gtrsp}) simplify to 
\begin{equation}
    \langle g \rangle = \frac{1}{2}.
\end{equation}
This result agrees with that obtained by Baranger and Mello \cite{baranger} and Brouwer and Beenakker \cite{brouwer} for a system with three ideal terminals in the dephasing regime.

\subsection{Intact Time-Reversal Symmetry}

Proceeding in the same way as above, we substitute Eqs. (\ref{gTRS}) and (\ref{g13}) in Eq. (\ref{Gdp}) and obtain that
\begin{eqnarray} \fl
    \langle g \rangle =
\frac{N_1 N_2}{N_1 + N_2} -\frac{N_1 N_2}{\left(N_1 + N_2\right)^2}\frac{1}{\left(1+\frac{\tau_d}{\tau_\phi}\right)}\nonumber \\ -\frac{N_1 N_2}{\left(N_1 + N_2\right)^2}\frac{\frac{\tau_d}{\tau_\phi}}{\left(1+\frac{\tau_d}{\tau_\phi}\right)^2}\gamma + \frac{N_1N_2}{N_1 + N_2}\frac{\left(\frac{\tau_d}{\tau_\phi}\right)^3}{\left( 1 + \frac{\tau_d}{\tau_\phi}\right)^3}\gamma^2 +  \mathcal{O}\left( \gamma^3\right), \label{gtrspp}
\end{eqnarray}
where \cite{baranger}
\be \frac{N_3}{N_1+N_2} = \frac{\tau_d}{\tau_\phi} .\ee In contrast with the broken TRS case, the Eq. (\ref{gtrspp}) depends on $\gamma$. The first term is in agreement with Baranger and Mello \cite{baranger}, while the second is the first correction due to the barrier in the terminal 3. Note that when $\tau_d\gg \tau_\phi$ ($N_3\gg N_1+N_2$) the quantum correction vanishes because of the dephasing effect, as expected. 

As done above, we can take that the limit $N_3\gg1$, while $N_1=N_2=1$, and Eq. (\ref{gtrspp}) simplifies to 
\begin{equation}
    \langle g \rangle = \frac{1}{2}-\frac{1}{2N_3}\sum_{k=0}^\infty \gamma^k = \frac{1}{2}-\frac{1}{2N_3(1-\gamma)}. \label{gN}
\end{equation}
When $\gamma=0$, we recover the result obtained by Brouwer and Beenakker, see Eq. (4.3) of \cite{brouwer}. 

\section{Conclusion}

We studied quantum transport through a system connected to three terminals having one of them as a tunnel barrier. Using a novel semiclassical approach based on a matrix integral representation, we obtained a general expression to the average of dimensionless transport moments Eq. (\ref{pT}) with broken TRS. This allows to calculation of any electronic transport moment as an expansion in power series of $\gamma$, even in the extreme quantum regime.  

We presented explicit expressions for four experimental observables: average conductance, conductance variance, average shot-noise power, and average third cumulant. We focused on three different architectures: symmetric terminals $N_1=N_2=N_3$; asymmetric terminals with $N_1=N_2$ and $N_3=1$; asymmetric terminals with $N_1=N_2=1$. We found that the average conductance is always close to asymptotic RMT results as a function of tunnel barrier for a small channel numbers, see Figs.\ref{moments_cases}.a and \ref{fig:my_label}.a. On the other hand, higher order cumulants are very sensitive to the number of channels when they are small; however, they converge to RMT as long as $M$ increases, see Figs. (\ref{moments_cases}.b-d) and (\ref{fig:my_label}.b-d). ). The difference between small $M$ and large $M$ for higher cumulants can be interpreted as a consequence of multifractality in transport dynamics because, as reported in Ref.\cite{nathan}, conductance fluctuations of chaotic billiard are multifractal in the former case and monofractal in the latter. Therefore, we can expect that multifractality is more prominent in higher cumulants than in average conductance. So the experimental measurement of this difference indirectly indicates multifractality or monofractality of transport dynamics.

We also calculated the average conductance of the system with intact TRS,  which allowed us to obtain expressions for the quantum interference correction of conductance in different limits. This is very sensitive to the value $M$, unlike average conductance. Therefore, the results only converge to RMT when the value of $M$ is significant, see Fig.\ref{WL}.

Furthermore, we revisited the dephasing model to show how the tunnel barrier can influence this effect. We generalized the results of  Ref. \cite{brouwer} including the tunneling barrier effect, Eq.\ref{gN}. 

\ack
Financial support from CAPES and from CNPq, grants 306765/2018-7 and 309457/2021-1, is gratefully acknowledged. We also thank Robert Whitney for some interesting discussions and two referees for their suggestions.

\section*{Appendix}

\subsection{Partitions}

A partition is a non increasing sequence of positive integers, and we say  $\mu \vdash n$ or $|\mu| = n$ if $\sum_i  \mu_i = n$. The number of parts is the length, $l(\mu)$. Partitions can be represented by a Young diagram, constructed by ordering boxes from left to right and from top to bottom, such that there is $\mu_i$ boxes in the $i$th row. We say that $\lambda$ is contained in $\mu$, $\lambda \subset \mu$, if $\lambda_i \leqslant \mu_i $ for all $i$, i.e. the Young diagram of $\lambda$ is covered by the Young diagram of $\mu$. For example, $(2,1) \subset (3,2,1)$. A skew diagram, $\mu \setminus \lambda$, is then defined to be the collection of those boxes in $\mu$ that do not belong to $\lambda$. When writing a partition, we use the notation $a^b$ to mean part $a$ appears $b$ times.

\begin{figure}
    \centering
    \includegraphics[scale = 0.5]{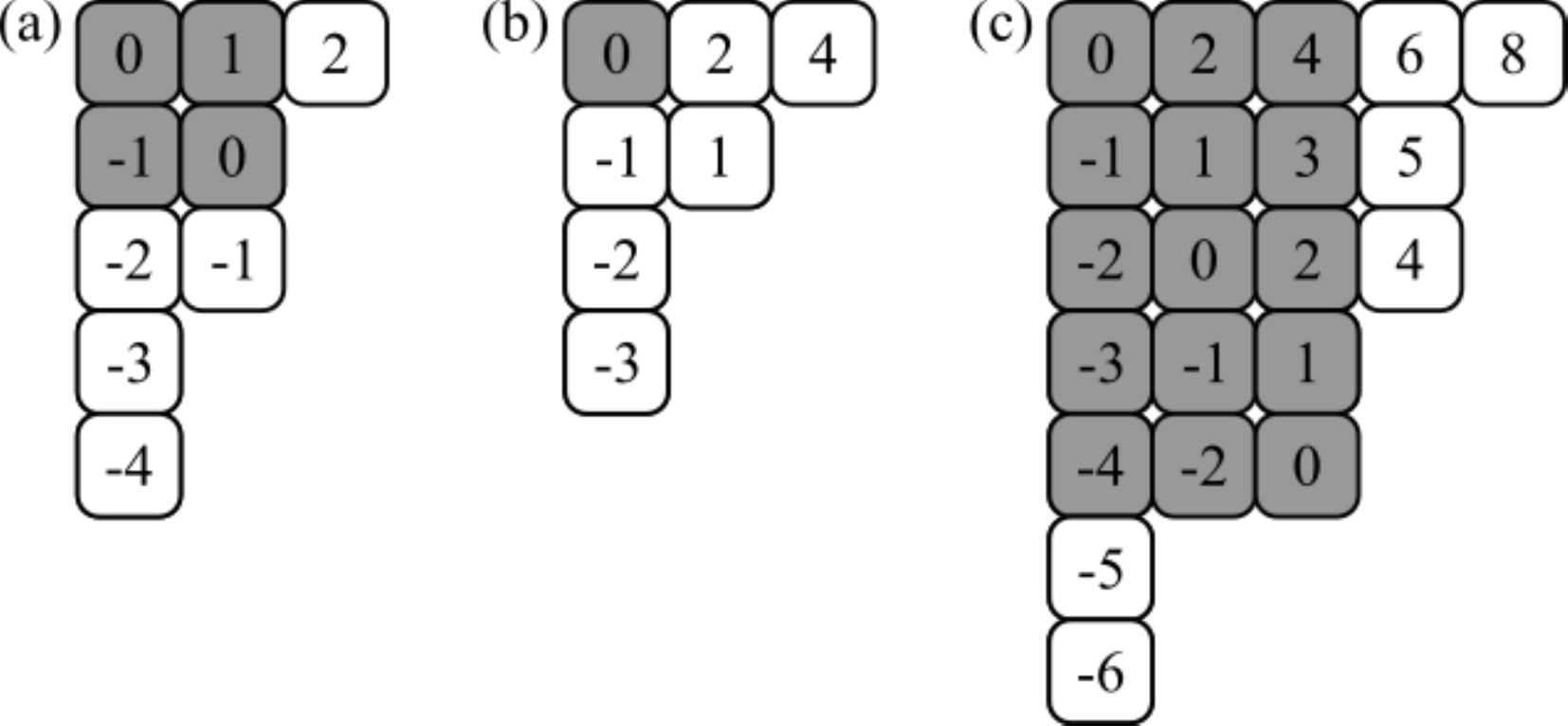}
    \caption{Young diagrams of three partitions: $(3,2^2, 1^2)$, $(3,2,1^2)$ and $(5,4^2, 3^2, 1^2)$. In (a) we show the 1-contents and the gray area is the Durfee square (or 1-Durfee rectangle), which has size 2. In (b) we have a double hook with its 2-contents; the trivial 2-Durfee rectangle is highlighted in gray. In (c) we show again the 2-contents, but in this case the 2-Durfee rectangle has width 3. As examples of skew-diagrams, we consider the white boxes. In (a), they correspond to $(3,2^2, 1^2)\setminus (2^2)$; in (b), to $(3,2,1^2)\setminus (1)$; in (c), to $(5,4^2, 3^2, 1^2)\setminus (3^5)$.}
    \label{partitions}
\end{figure}

If a box occupies the $j$th position in the $i$th row, is has coordinates $(i,j)$. Given the coordinate of a box, its $\alpha$-content is defined as $c_\alpha(i,j) = \alpha(j - 1) - i + 1$. The $\alpha$-Durfee rectangle is the smallest rectangle in the Young diagram that contains all zero $\alpha$-contents. For example, the partition $(3,2,1^2)$ has a 2-Durfee rectangle of width $1$ while for $(5,4^2,3^2,1^2)$ the width of the 2-Durfee rectangle is $3$. When $\alpha = 1$, $D_\alpha(\mu)$ is also the size of the biggest square contained in the diagram of $\mu$. Some examples are shown in Fig.~\ref{partitions}. Partitions of the form $(m, 1^{k})$, are called hooks, while partitions of the form $(p, q, 1^{r})$, we call double-hooks (of course, all hooks are double-hooks). For any hook we have $D_1(\mu)=1$, while for any double-hook $D_2(\mu)=1$.

The product of all non zero contents is denoted as
\begin{equation}
    t_{\alpha}(\mu) = \prod_{i = 1}^{l(\mu)}\prod_{\stackrel[c_\alpha \neq 0]{}{j = 1}}^{\mu_i}c_\alpha(i,j).
\end{equation}
Using the $\alpha$-content we are able to define a family of monic polynomials,
\begin{equation}\label{stoN}
    \left[x\right]^{(\alpha)}_\mu = \prod_{i = 1}^{l(\mu)}\prod_{j = 1}^{\mu_i}\left( x + c_\alpha(i,j) \right).
\end{equation} 
The importance of $D_\alpha(\mu)$ and $t_{\alpha}(\mu)$ for us is that, for small $x$, we have
\begin{equation}\label{expanN}
\left[x\right]^{(\alpha)}_\mu= t_{\alpha}(\mu) x^{D_\alpha(\mu)} + \mathcal{O}\left(x^{D_\alpha(\mu)+1}\right).
\end{equation}

Let $S_n$ be the group of all permutations acting on the set $\{ 1, 2, \cdots, n \}$. The irreducible representations and the conjugacy classes of $S_n$ are labelled by partitions. The dimension of the irrep labelled by $\mu$ is given by
\begin{equation}
    d_\mu = n! \prod_{i = 1}^{l(\mu)}\frac{1}{\left( \mu_i - i + l(\mu)\right)!}\prod_{j = i + 1}^{\mu_i} \left(\mu_i - \mu_j + j - i\right).
\end{equation}
When $\mu\vdash m$ is a hook, $\mu = (m - k, 1^{k})$, its dimension is $d_\mu={m - 1\choose k}$. 

The irreducible character of a conjugacy class $\lambda$ in the irreducible representation $\mu$ is denoted $\chi_\mu(\lambda)$. They obey the following orthogonality relation
\begin{equation}
    \sum_{\pi \in S_n}\chi_\mu(\pi) \chi_\lambda(\pi \xi) = \frac{n!}{d_\lambda}\delta_{\mu\lambda}\chi_\lambda(\xi).
\end{equation}
Note that $\chi_\mu(1^{|\mu|})=d_\mu$.

\subsection{Symmetric polynomials}

Polynomial irreducible representations of the unitary group $\mathcal{U}(N)$ are also labelled by partitions $\mu\vdash n$, with length smaller or equal to $N$. Schur polynomials are the characters of these representations, orthogonal with respect to integration against the invariant measure. They can be obtained by
\begin{equation}\label{schurdef}
    s_\mu(X) = \frac{\det \left( x_i^{n + \mu_i - i}\right)}{\Delta (X)},
\end{equation}
where $\mu \vdash n$, $\Delta(X)$ is the Vandermonde determinant and $l(\mu) \leqslant N$ (if $l(\mu) > N$ then $s_\mu(X) = 0$). Their value at the identity is related to some polynomials defined previously, 
\begin{equation}
    s_\mu\left( 1^N\right) = \frac{d_\mu}{n!}\left[N\right]_\mu^{(1)}.
\end{equation}

Schur polynomials form a basis for the ring of symmetric functions \cite{macdonald}. Their product can be expressed as 
\begin{equation}\label{lrcoef}
s_\mu(X)s_\rho(X) = \sum_{\alpha}C^{(1)}_{\mu\rho\alpha}s_\alpha(X).
\end{equation}
The coefficient in these expansion are called Littlewood-Richardson coefficients. They are null unless $\rho \subset \alpha $, $\mu \subset \alpha$ and $|\alpha| = |\mu| + |\rho|$. In general, there is no closed form for these coefficients, but there are some rules to obtain them. 

In terms of the Littlewood-Richardson coefficients, so-called skew Schur polynomials are defined as
\begin{equation}
    s_{\alpha\setminus \mu}(X) = \sum_\rho C^{(1)}_{\mu\rho\alpha}s_\rho(X),
\end{equation}
where the sum runs over all partitions such that $\rho \subset \alpha $, $\mu \subset \alpha$ and $|\alpha| = |\mu| + |\rho|$. When $X=1^N$, this reduces to
\begin{equation}\label{skewschur1}
    s_{\mu\setminus \lambda}\left(1^N\right) = \det \left( {N + \mu_i - i - \lambda_i + j - 1\choose \mu_i - i - \lambda_i + j} \right).
\end{equation}
If $\mu$ is a hook and $\lambda=(1)$ then
\begin{equation}\label{schurespecial}
s_{(m + 1 - k, 1^k)\setminus (1)}(1^{N}) = \frac{(N)^{m-k}(N)_{k}}{(m-k)!k!},
\end{equation}
where $(N)^{m-k}$ and $(N)_{k}$ are rising and falling factorials.

Another useful relation is the Cauchy identity, where a power of a determinant is expressed through Schur polynomials as
\begin{equation}\label{cauchyschur}
\det(1-\gamma X)^{-N} = \sum_{\rho}s_\rho\left(\gamma 1^{N}\right)s_\rho(X).
\end{equation}
This infinite sum runs over all partitions of all integers, like a generalization of the geometric series. 

Another special family of symmetric functions are the zonal polynomials, which we denote $Z_\mu(X)$. They are not irreducible characters, but they are orthogonal with respect to an integral over the space of symmetric unitary matrices \cite{macdonald}. In contrast to the Schur polynomials, they have no explicit expression like Eq.~(\ref{schurdef}). They are also related to the polynomials defined previously, $Z_\mu(1^N) = \left[ N\right]_\mu^{(2)}$. 

Zonal polynomials also form a basis for the ring of symmetric functions \cite{macdonald}. Products can be expressed as linear combinations, with different Littlewood-Richardson coefficients 
\begin{equation}\label{lrcoefzonal}
Z_\mu(X)Z_\rho(X) = \sum_{\alpha}C^{(2)}_{\mu\rho\alpha}Z_\alpha(X).
\end{equation}
Again, the coefficients vanish unless $\rho \subset \alpha $, $\mu \subset \alpha$ and $|\alpha| = |\mu| + |\rho|$. Another useful expansion using zonal polynomials is 
\begin{equation}
\mathrm{Tr}\left (X^{k+1}\right )= \frac{1}{(2k+1)!} \sum_{\lambda \vdash k+1} d_{2\lambda} t^{(2)}_\lambda Z_\lambda(X).
\end{equation}
We also have an analogous to Cauchy identity to zonal polynomials
\begin{equation}\label{cauchyzonal}
\det{(1 -\gamma X)}^{-\frac{N}{2}}  = \sum_\rho \frac{1}{j_\rho}Z_\rho(\gamma 1^{N})Z_\rho(X),
\end{equation}
with $j_\rho = \left|2\rho\right|!/d_{2\rho}$, i.e., the product of hook-lengths in the partition $2\rho$ \cite{stanley}. 

\subsection{Selberg-Jack integrals}

The Selberg integral is
\begin{equation}
    S_N(a,b,c)=\int_0^1 \det\left(X^{a-1}(1-X)^{b-1}\right)|\Delta(X)|^{2c}dX,
\end{equation}
where $X$ is of size $N$, and it is given by
\begin{equation}
    S_N(a,b,c)=\prod_{j=1}^{N}\frac{\Gamma(a+(j-1)c)\Gamma(b+(j-1)c)\Gamma(1+jc)}{\Gamma(a+b+(N+j-2)c)\Gamma(1+c)}.
\end{equation}
The simplest case $N=1$ is the Beta integral,
\begin{equation}
    S_1(a,b,0)=\int_0^1 x^{a-1}(1-x)^{b-1}dx=\frac{\Gamma(a)\Gamma(b)}{\Gamma(a+b)}.
\end{equation}

If $c=a$, then by taking $X=\frac{Yd}{b}$ and letting $b\to\infty$ we get as a consequence the integral
\begin{equation}\label{SelbExp}
    \int_0^\infty e^{-d\mathrm{Tr}Y}\det(Y)^{a-1}|\Delta(Y)|^{2a}dY=\frac{1}{d^{aN^2}}\prod_{j=1}^{N}\frac{\Gamma(ja)\Gamma(1+ja)}{\Gamma(1+a)}.
\end{equation} 
On the other hand, taking $X=\frac{1}{2}-\frac{Y}{2\sqrt{2b}}$ and letting $a=b\to \infty$ we get the Mehta integral,
\begin{equation}
    \int_0^\infty e^{-\mathrm{Tr}Y^2/2}|\Delta(Y)|^{2c}dY=(2\pi)^{N/2}\prod_{j=1}^{N}\frac{\Gamma(1+jc)}{\Gamma(1+c)}.
\end{equation}

The Selberg integral can be generalized to the Selberg-Jack integral \cite{kaneko,kadell}, 
\begin{equation}\label{jackselberg}
    SJ_N(a,b,\alpha)=\int_0^1 J_\lambda^{(\alpha)}(X)\det\left(X^{a-1}(1-X)^{b-1}\right)|\Delta(X)|^{2/\alpha}dX,
\end{equation}
where $J_\lambda^{(\alpha)}(X)$ are the Jack polynomials \cite{stanley}. This integral is given by
\begin{equation}\fl
    SJ_N(a,b,\alpha)=J_\lambda^{(\alpha)}(1^N)\prod_{j=1}^{N}\frac{\Gamma(\lambda_j+a+(N-j)/\alpha)\Gamma(b+(N-j)/\alpha)\Gamma(1+j/\alpha)}{\Gamma(\lambda_j+a+b+(2N-j-1)/\alpha)\Gamma(1+1/\alpha)}.
\end{equation} This is relevant to us because When $\alpha=1$ the Jack polynomial is proportional to the Schur polynomial, $J_\lambda^{(1)}(X)=\frac{|\lambda|!}{d_\lambda}s_\lambda(X)$, while for $\alpha=2$ it equals the zonal polynomial, $J_\lambda^{(2)}(X)=Z_\lambda(X)$.

\end{document}